\documentclass[12pt]{iopart}
\usepackage{graphicx}
\usepackage{epstopdf}

\newcommand{\Lg}{L_{\rm geo}}
\newcommand{\Lj}{L_{\rm j}}

\newcommand{\Peo}{\Phi_{\rm e0}}

\begin{document}

\title[]{Protecting SQUID metamaterials against stray magnetic field}

\author{S Butz$^1$, P Jung$^1$, L V Filippenko$^{2,3}$, V P Koshelets$^{2,3}$ and A~ V~ Ustinov$^{1,3}$}

\address{$^1$ Physikalisches Institut, Karlsruhe Institute of Technology,
76131 Karlsruhe, Germany}

\address{$^2$ Kotel'nikov Institute of Radio Engineering and Electronics (IREE RAS), Moscow 125009, Russia}

\address{$^3$ National University of Science and Technology MISIS, Moscow 119049, Russia}

\ead{s.butz@kit.edu}

\begin{abstract}
Using superconducting quantum interference devices (SQUIDs) as basic, low-loss elements of thin-film metamaterials has one main advantage: Their resonance frequency is easily tunable by applying a weak magnetic field. The downside, however, is a strong sensitivity to stray and inhomogeneous magnetic fields. In this work, we demonstrate that even small magnetic fields from electronic components destroy the collective, resonant behaviour of the SQUID metamaterial. We also show how the effect of these fields can be minimized. As a first step, magnetic shielding decreases any initially present fields including the earth's magnetic field. However, further measures like improvements in the sample geometry have to be taken to avoid the trapping of Abrikosov vortices. 
\end{abstract}

\maketitle

\section{Introduction}
The field of superconducting metamaterials that employ the nonlinear inductance of a Josephson junction as tunable element is just emerging. Unlike other superconducting metamaterials \cite{ricci2007, gu2010}, the tuning of the so-called Josephson inductance does not degrade the quality factor of the resonance over almost the full range of tunability.
The idea of a metamaterial consisting of superconducting quantum interference devices (SQUIDs) was theoretically introduced and investigated in \cite{lazarides2007, gabitov2012, du2008}. Recently, the single junction (rf-) SQUID as magnetic field tunable meta-atom has been experimentally investigated in \cite{jung13} and employed successfully as basic building block of a metamaterial in \cite{butz13}. However, the easily accessible, broad range tunability comes at the cost of a high sensitivity to external magnetic fields. 

Here, we demonstrate and discuss measures that are necessary to use rf-SQUIDs as building blocks of a superconducting metamaterial. In order to generate a coherent response, all SQUIDs have to have the same resonance frequency at the same magnetic field, i.e. the resonance curves of the individual SQUIDs have to overlap.

Stray magnetic fields are either caused by magnetic components used in the experimental setup or outside sources such as the earth's magnetic field. They have two main effects that disturb the collective behaviour. First, a spatially inhomogeneous field is created at the sample. This means that each SQUID is biased with a different magnetic field. Second, if there is a magnetic field present when cooling the sample from above to below the critical temperature of the superconductor, Abrikosov vortices \cite{abrikosov1957} can be trapped in the thin Nb film. The vortices create local magnetic fields that remain in the superconductor until the sample is warmed up again.

Thus, any stray magnetic field has to be screened as well as possible and, additionally, the sample has to be designed in such a way as to discourage the trapping of Abrikosov vortices. The latter can be done by using normal metal instead of superconductors where possible and by decreasing the width of the superconducting structures in order to reduce their demagnetization factor \cite{stan04, polyakov2007}. Additionally, these vortices are trapped preferably at inhomogeneities in the superconductor \cite{polyakov2007}, for example at vias where two layers of superconductor connect. Thus, the area of these vias should be chosen to be as small as possible.

\section{The SQUID metamaterial}
\begin{figure}[htbp]
 \centering
 \includegraphics[width = 1\textwidth]{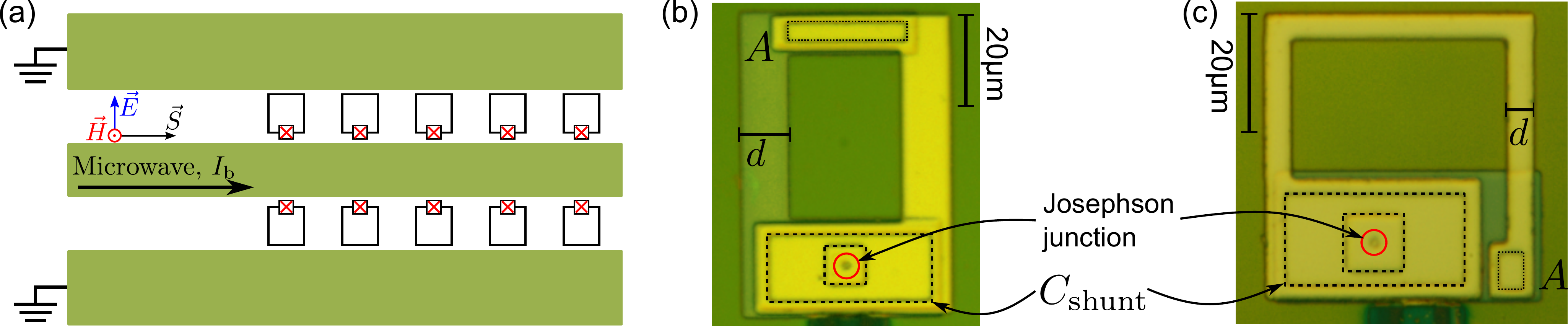}
 \caption{(a) Coplanar waveguide geometry including the rf-SQUIDs. The vectors $\vec{E}$, $\vec{H}$ and $\vec{S}$ denote the direction of electric field, magnetic field and Poynting vector, respectively (b) Single rf-SQUID used in sample S1 including junction, shunt capacitance $C_{\rm shunt}$, via area $A$ and line width $d$ as mentioned in the text. (c) Single rf-SQUID used in sample S2, note the different line width $d_{\rm S1} = 10\,\mu$m versus $d_{\rm S2} = 4\,\mu$m and via area $A_{\rm S1} = 25 \times 3\,\mu \rm m^2$ versus $A_{\rm S2} = 5 \times 3\,\mu \rm m^2$.}
 \label{fig:waveguide}
\end{figure}

We performed experiments with a one-dimensional metamaterial that contains rf-SQUIDs as meta-atoms. A chain of these rf-SQUIDs is placed inside each gap of a coplanar waveguide (CPW) (see figure\,\ref{fig:waveguide}(a)). The SQUIDs as well as the CPW are fabricated on a Si substrate using a $\rm Nb$/$\rm AlO_x$/$\rm Nb$ trilayer process. A small amplitude microwave signal travels along the waveguide. In addition, the central conductor is used to bias the rf-SQUIDs with a dc magnetic field by applying a dc current $I_b$. Note, that in the presentation of the measured data in section\,\ref{sec:meas}, instead of magnetic field we use the more natural quantity magnetic flux per SQUID loop. 

In this work, we consider two different samples, each containing 54~SQUIDs, 27~per gap. Sample S1 ccontains a fully superconducting waveguide made of Nb. One of the SQUIDs, used in sample S1, is shown in figure\,\ref{fig:waveguide}(b). Its parameters, namely critical current of the Josephson junction $I_c$, the resulting zero field Josephson inductance $\Lj$ as well as the geometric inductance of the SQUID loop, are given in table\,\ref{tab:parameters}. 

\noindent The rf-SQUID of the second sample S2 is shown in figure\,\ref{fig:waveguide}(c). Its parameters are different from those of the SQUIDs of sample S1. They are also given in table\,\ref{tab:parameters}. Both SQUIDs have a $\beta_{\rm L} = 2\pi\Lg I_c/\Phi_0  < 1$ and use a parallel plate capacitor $C_{\rm shunt}$ that shunts the junction to increase the total capacitance $C_{\rm tot}$ of the circuit, thus decreasing the resonance frequency.  

Due to the different design, the occurrence of Abrikosov vortices should be considerably suppressed in sample S2. For example, the SQUIDs used in S2 have a smaller width $d$ of the superconducting leads and area of the via $A$. Additionally, the ground planes of the CPW of sample S2 are made of normal metal (Pd) instead of Nb. The central conductor is still made from Nb, due to requirements of the fabrication process.

\begin{table}[btp]
\caption{\label{tab:parameters} SQUID parameters for the two different samples S1 and S2 used in the measurements shown in section\,\ref{sec:meas}.}
\begin{indented}

 \item[]\begin{tabular}{@{}llllll}
 \br
 & $I_c$ [$\mu$A]& $\Lj (\Peo = 0)$ [pH] & $\Lg$ [pH]&$\beta_L$& $C_{\rm tot}$ [pF]\\
 \mr
 S1 &3.4&97&78&0.80&1.5\\
 S2& 1.8 &183&83&0.45 &2.0\\
 \br
\end{tabular}
\end{indented}

\end{table}


\section{Measurement Setup}
The $4\rm mm \times 4 mm$ Si chip that holds the CPW with the SQUIDs is glued to a printed circuit board that is used to connect the coaxial cables to the CPW. This setup is then installed inside a cylindrical copper sample holder. The sample holder, including the microwave electronics, i.e. attenuators, circulator, amplifier and bias tees, is placed inside a cylindrical cryoperm shield to suppress external magnetic fields. The bias tees are used to superpose the microwave signal with the dc current used to tune the resonance frequency of the rf-SQUIDs.

\begin{figure}[htbp]
 \centering
 \includegraphics[width = .45\textwidth]{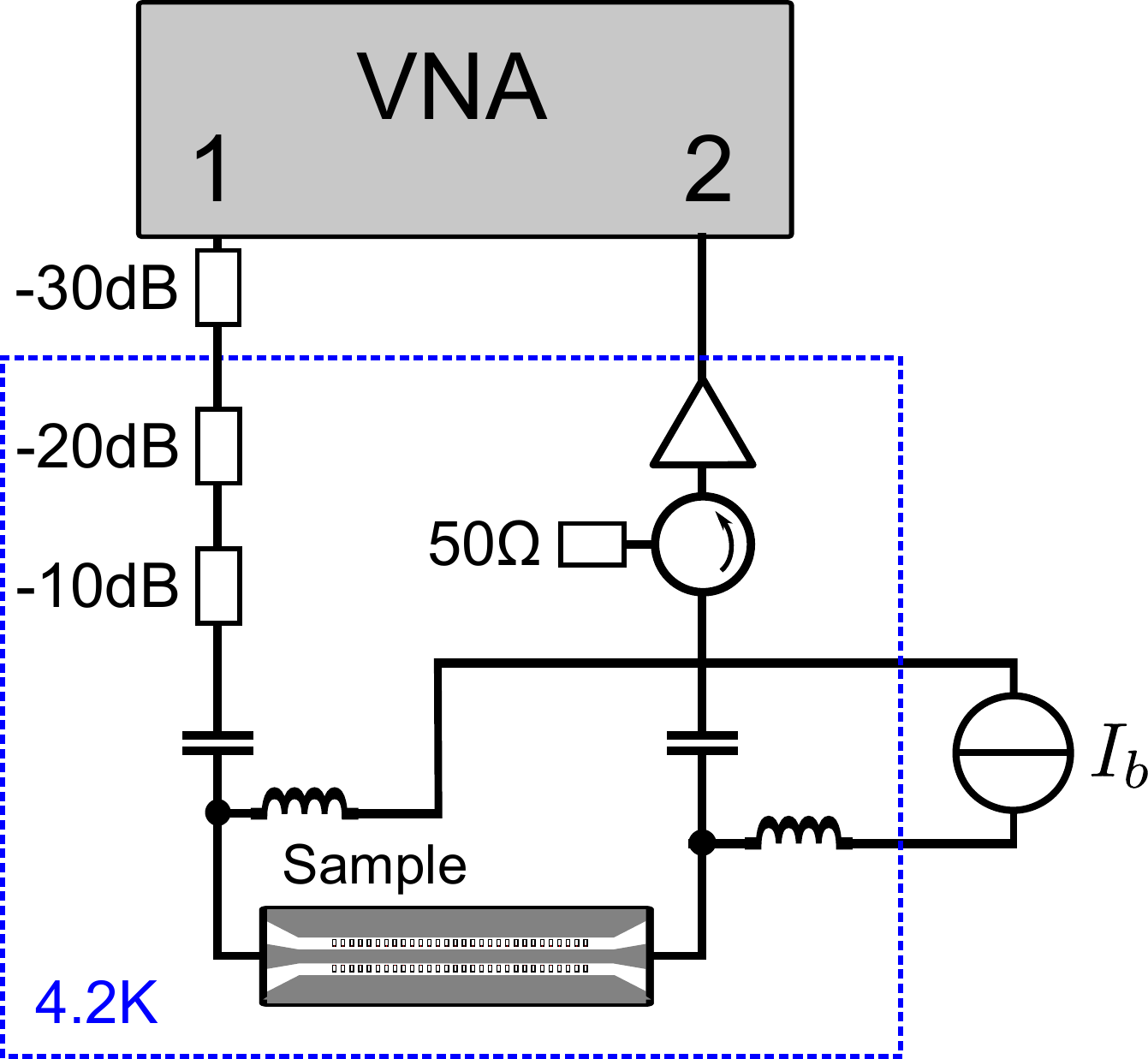}
 \caption{Measurement setup including the vector network analyzer (VNA), bias tees, attenuation, circulator and cryogenic amplifier.}
 \label{fig:setup}
\end{figure}

The setup is then installed on a dip stick and immersed in liquid helium. The microwave lines are connected to a vector network analyzer (VNA) using additional warm attenuation at the input. The full setup, apart from the cryoperm shield, is shown in figure\,\ref{fig:setup}.

\noindent We measure the field and frequency dependent complex transmission $S_{21}$ through the sample.

\section{Experimental Results}\label{sec:meas}

\begin{figure}[htbp]
 \centering
 \includegraphics[width = .45\textwidth]{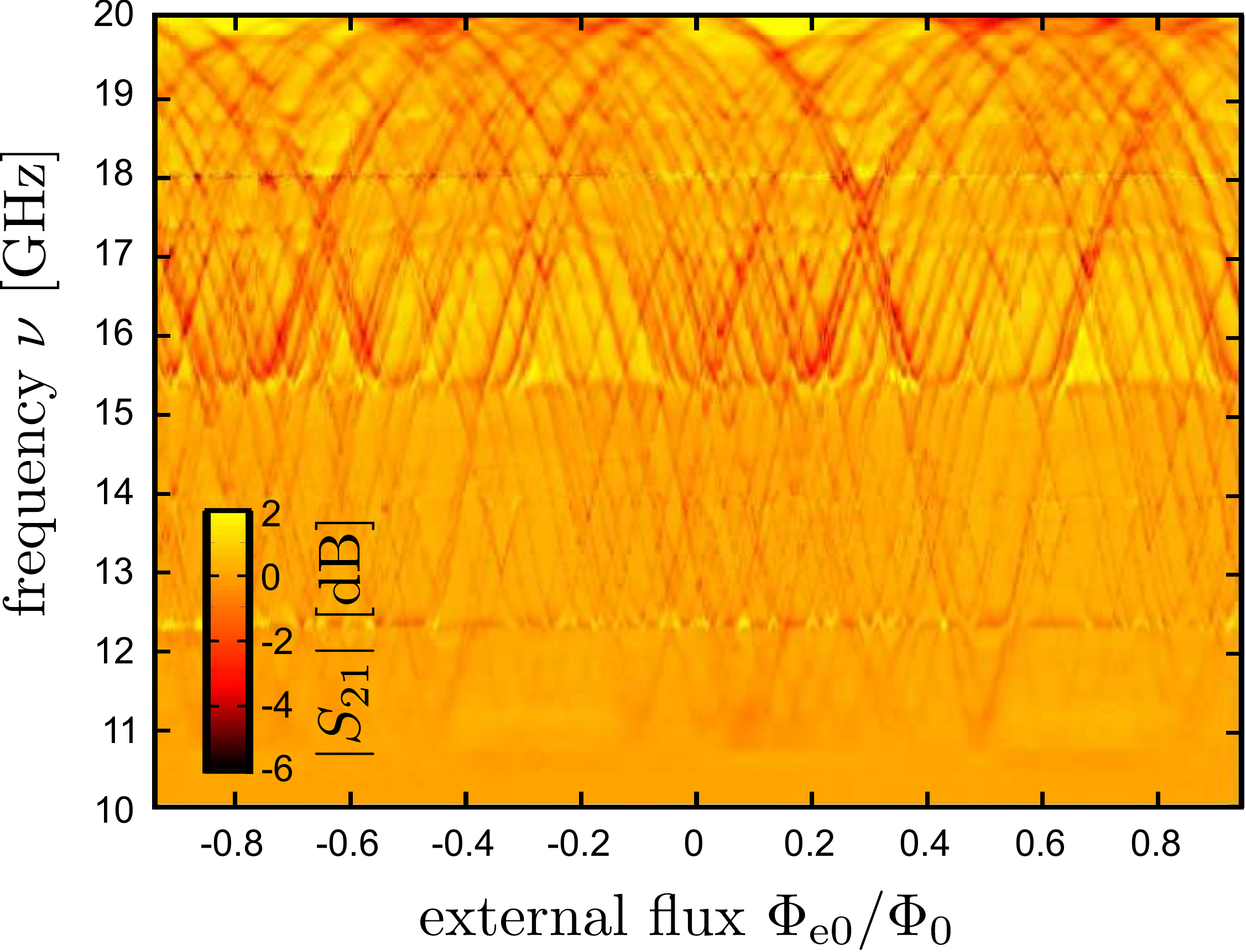}
 \caption{Flux and frequency dependent transmission magnitude measured on sample S1. The measured data is normalized along the flux axis in order to improve visibility of the lines.}
 \label{fig:manylines}
\end{figure}

\begin{figure}[htbp]
 \centering
 \includegraphics[width = .45\textwidth]{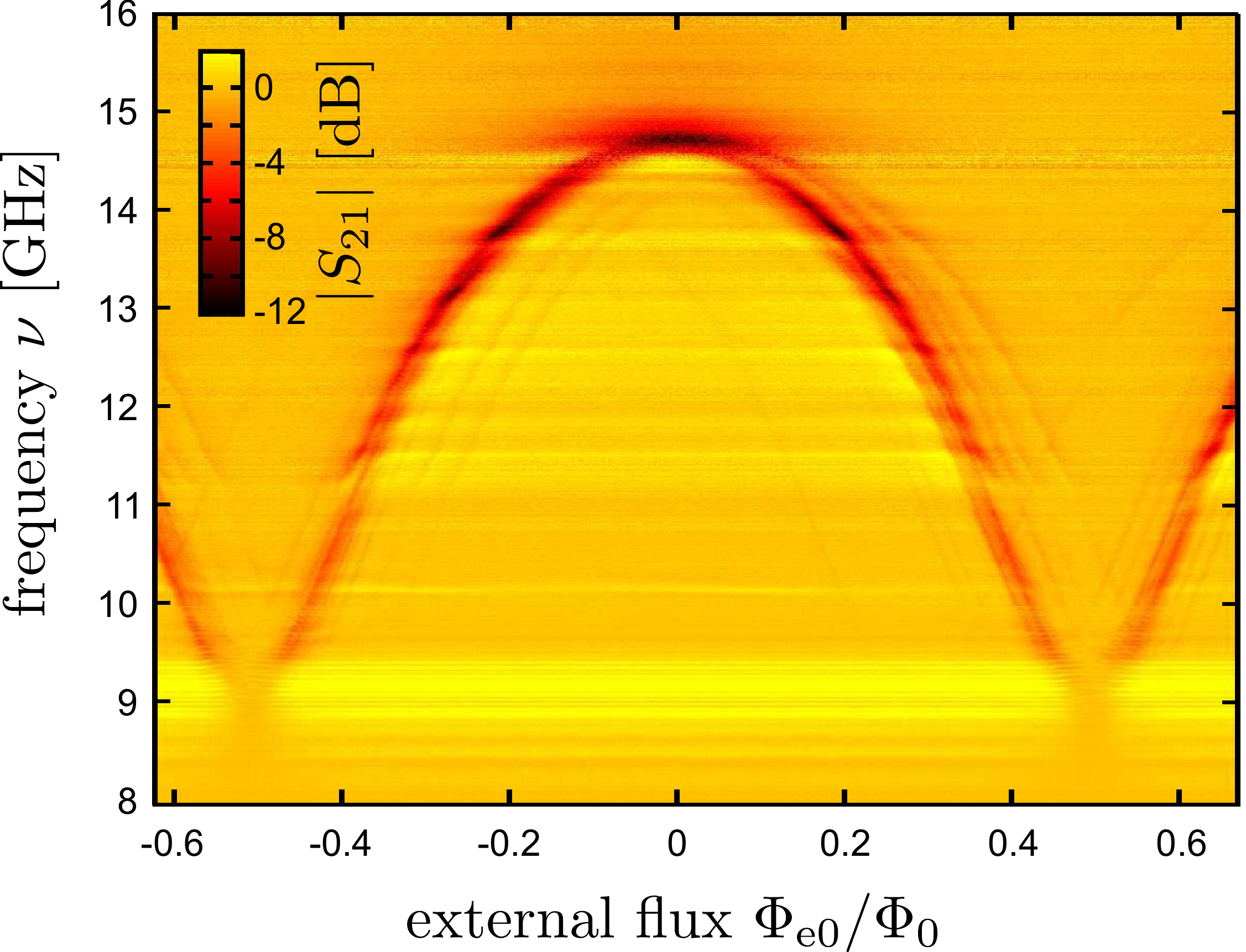}
 \caption{Transmission magnitude through sample S2 when the microwave cable does not pass the sample. Note the the calibration for this measurement was done in situ, using the through calibration function of the VNA at an external flux $\Peo = 0.5\Phi_0$.}
 \label{fig:nospread}
\end{figure}

Here, we present experimental results obtained with different setups and different samples S1 and S2. First, we consider a measurement performed on sample S1. For better visibility, all measurements except the one shown in figure\,\ref{fig:nospread}, are normalized for each frequency value with the average transmission magnitude along the flux axis. The resulting transmission data is presented colour coded in figure\,\ref{fig:manylines}. The picture shows many lines spread randomly over the full flux range.

Every line corresponds to the magnetic field dependent resonance curve of one or a small number of rf-SQUIDs and each of these individual or small groups of SQUIDs is thus biased with different magnetic field. The lines at approximately 11\,GHz, 12.2\,GHz and 15.5\,GHz are internal parasitical sample holder resonances that couple to the SQUIDs and distort the resonance lines. When looking more closely, one notices that different individual lines seem to have a slightly different periodicity in magnetic flux as well as a slightly different frequency range. This is an effect of a small spread in SQUID parameters.

In order to decrease the spread in magnetic flux bias, sample S1 was replaced by sample S2 which is built to reduce the trapping of Abrikosov vortices. However, the resulting transmission again shows the same forest of lines (not shown here). Thus, further measures to improve the setup had to be taken.

\begin{figure}[htbp]
 \centering
 \includegraphics[width = .45\textwidth]{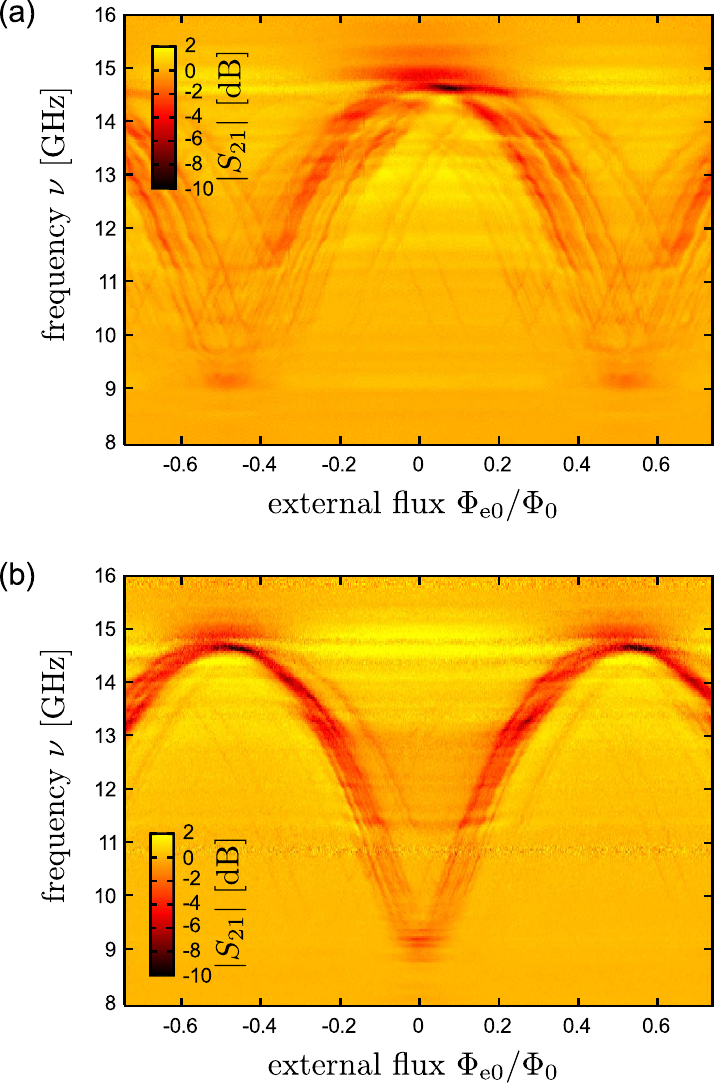}
 \caption{Transmission magnitude through sample S2. (a) All necessary microwave electronics, such as the amplifier and the bias tees, are installed outside the magnetic shield. However, a microwave cable is close to the top of the sample and the sample is placed in the upper third of the shield. (b) Sample is moved deep into the shield (lower third) and the microwave cable passes at the side of the sample.}
 \label{fig:effectcables}
\end{figure}

\noindent First, all components were examined to determine if and how strongly they are magnetic. The strongest magnetic field is created by the circulator. Since we used it only as protection from reflections from the amplifier back to the sample, we replaced it by a $3$\,dB attenuator which serves the same purpose. The amplifier and the bias tees are also magnetic. They, however, are essential to our measurement setup. In order to protect the sample from their fields, they a placed outside the cryoperm shield. In figure\,\ref{fig:effectcables}(a), it is clearly visible that these measures improved the behaviour of our SQUID metamaterial significantly. The spread of the lines is strongly reduced. Note, that due to the different SQUIDs parameters, the frequency band in which the resonance frequency is tunable is changed according to the values given for S2 in table\,\ref{tab:parameters}.
Yet, the microwave cables are also magnetic. This creates another complication. They have to connect to the sample and cannot be removed or installed far away. Thus, care has to be taken that they do not pass closely by the sample. The cables that are used close to the sample contain a central conductor made of copper and silver plated steel and an outer conductor made of tin plated copper. Figure\,\ref{fig:effectcables}(a),(b) and figure\,\ref{fig:nospread} show how the magnetic environment is considerably improved by careful arrangement of the cables. In figure\,\ref{fig:effectcables}(a) one cable passes on top of the sample holder with a distance of approximately $10$\,mm to the sample. In (b) this cable is moved to pass along the side of the sample and the sample is moved as far into the cylindrical cryoperm shield as possible. In figure\,\ref{fig:nospread}, a different sample holder at the same position deep inside the cryoperm shield is used, which allows to lead the microwave cables away from the 
sample without passing it. Finally, with this improved setup, a main resonance is clearly visible. Figures\,\ref{fig:effectcables} and \ref{fig:nospread} show that 
protecting the sample from magnetic fields is crucial, even the stray fields of microwave cables affect the outcome. It should be emphasized, that this result was obtained with sample S2.

\begin{figure}[htbp]
 \centering
 \includegraphics[width = .45\textwidth]{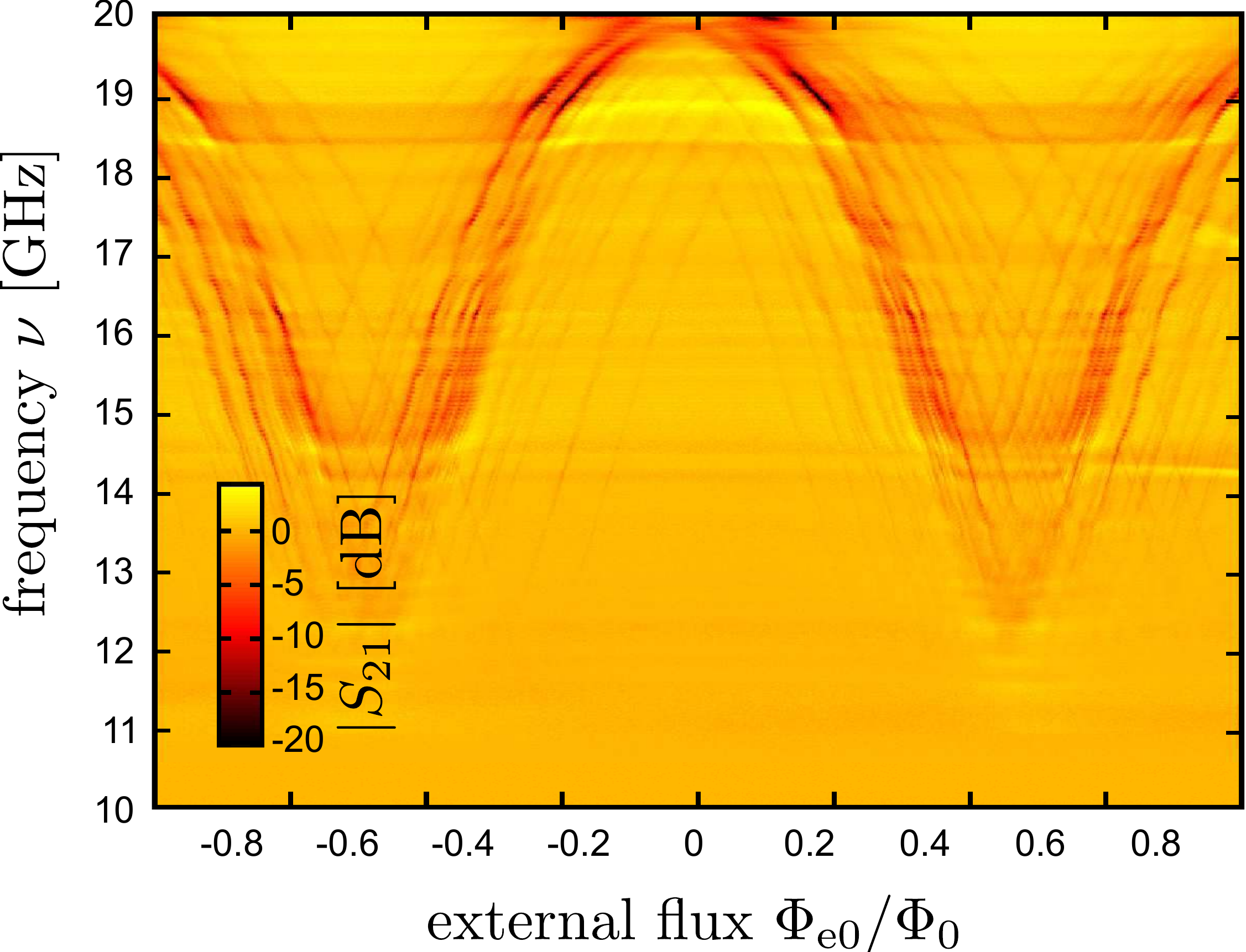}
 \caption{Transmission magnitude through sample S1 with the same setup used in the measurement shown in figure\,\ref{fig:nospread}.}
 \label{fig:meass1_2}
\end{figure}

We now change back to sample design S1. When we install it in exactly the same setup as the one used for the measurement on sample S2 in figure\,\ref{fig:nospread}, the quality of the result is degraded again (see figure\,\ref{fig:meass1_2}). The deviation in critical current as well as capacitance is less than $6\%$ for both samples. However, as mentioned above, there is also some spread in the periodicity in magnetic flux. This is due to slightly different areas of the individual SQUIDs, since the magnetic bias is applied along along the central conductor and should be homogeneous. For sample S1, the spread is up to $12\%$, while it is less (about $2\%$) for sample S2. This deviation contributes to the degradation, since the resonance curves of deviating SQUIDs are shifted by up to $0.1\Phi_0$ against the standard curves in the flux interval shown in figure\,\ref{fig:meass1_2}. However, this 
deviation is neither strong enough to explain the full spread in magnetic flux nor does it affect all resonance curves. Instead, the main difference between the two samples S1 and S2 is their affinity to trap Abrikosov vortices. Thus, the vortices are most probably the reason for the degradation of the performance of sample S1 under otherwise identical measurement conditions as for sample S2.

\section{Conclusion}
On the road towards creating a tunable magnetic metamaterial with a collective resonance frequency, we had to solve a few challenges by overcoming the effects of stray magnetic fields. We have shown that the inhomogeneous magnetic flux bias of the individual SQUIDs arises from fields due to magnetic components of the measurement setup as well as trapped Abrikosov vortices in the superconducting film. Apart from shielding the sample magnetically, two main measures have to be taken to protect against both effects. First, magnetic components that cannot be omitted have to be placed as far away from the sample as possible. Care has to be taken even with coaxial cables. Second, the trapping of Abrikosov vortices has to be prevented by using superconducting planes only where necessary and by decreasing the width of the superconducting leads of the SQUID and the area of the via which creates the contact between different superconducting layers. Applying all these measures, we were able to achieve a collective, 
tunable resonance curve of almost all 54 rf-SQUIDs.

\ack
The authors would like to acknowledge interesting and productive discussions with S. M. Anlage, I. Gabitov, N. Lazarides and G. Tsironis. This work was supported in part by the Ministry of Education and Science of the Russian Federation, by the Russian Foundation of Basic Research, and also by the Deutsche Forschungsgemeinschaft (DFG) and the State of Baden-W\"urttemberg through the DFG Center for Functional Nanostructures (CFN). Philipp Jung would like to acknowledge the financial support by the Helmholtz International Research School for Teratronics (HIRST), Susanne Butz would like to acknowledge the financial support by the Landesgraduiertenf\"orderung Baden-W\"urttemberg.

\end{document}